\documentclass[journal]{IEEEtran}
\makeatletter
\def\subsubsection{
  \@startsection
    {subsubsection}
    {3}
    {\parindent}
    {3.5ex plus 1.5ex minus 1.5ex}
    {0.7ex plus .5ex minus 0ex}     
    {\normalfont\normalsize\itshape}
}
\makeatother
\usepackage{amsmath,amsfonts}
\usepackage{algorithmic}
\usepackage{algorithm}
\usepackage{array}
\usepackage[caption=false,font=footnotesize]{subfig}
\usepackage{textcomp}
\usepackage{stfloats}
\usepackage{url}
\usepackage{verbatim}
\usepackage{graphicx}
\usepackage{cite}
\usepackage{hyperref}
\usepackage{url}
\usepackage{cleveref}
\usepackage{siunitx}
\usepackage[export]{adjustbox}
\usepackage{tikz}
\usepackage{makecell}
\usetikzlibrary{arrows.meta,patterns,calc,decorations.pathreplacing}

\begin{document}

\title{Efficient Multi-Market Scheduling of Virtual Power Plants via Spectral Representation of Uncertainty}

\author{Lorenzo~Zapparoli,~\IEEEmembership{Graduate Student Member,~IEEE},
        Blazhe~Gjorgiev,
        Giovanni~Sansavini,~\IEEEmembership{Member,~IEEE}
\thanks{Lorenzo Zapparoli, Blazhe Gjorgiev, and Giovanni Sansavini are with the Institute of Energy and Process Engineering, ETH Zurich, Zurich, Switzerland (e-mail: lzapparoli@ethz.ch, gblazhe@ethz.ch, sansavig@ethz.ch)}}

\maketitle
\begin{abstract}
As the penetration of distributed energy resources increases, harnessing their flexibility becomes critical for power system operations. Virtual power plants (VPPs) offer a promising solution. However, existing VPP market scheduling tools exhibit a tradeoff between economic performance and tractability. Stochastic formulations provide probabilistically optimal decisions but are computationally intractable for large systems due to scenario explosion. Robust approaches are more tractable but often yield conservative decisions.
This paper addresses this gap by proposing a stochastic multi-market VPP scheduling framework that represents uncertainty in the spectral domain via intrusive Polynomial Chaos Expansion (PCE). The resulting reformulation yields a low-dimensional deterministic spectral counterpart that preserves the stochastic structure and can be solved efficiently with standard optimization tools.
The proposed spectral approach is demonstrated on a DER-based VPP operating on a realistic Swiss low-voltage grid and benchmarked against a state-of-the-art scenario-based solution. Results show that intrusive PCE achieves solution quality comparable to the scenario-based benchmark, with up to a 137 times reduction in computational effort, while yielding highly accurate bidding decisions. Finally, to facilitate adoption and reproducibility, we release an open-source, application-agnostic projection tool that automates the spectral reformulation for generic single- and two-stage stochastic programs.
\end{abstract}


\section{Introduction}
\label{intro}
\IEEEPARstart{T}{he} transition toward a low-carbon energy system is transforming the way electricity is generated, consumed, and transmitted. The growing integration of renewable energy sources (RES)~\cite{IRENA2022} and the rapid deployment of distributed energy resources (DERs)~\cite{IEA_Report} present both opportunities and challenges for grid operators. While these technologies support decarbonization efforts, their variability and decentralization increase the demand for system flexibility~\cite{IEA_Report}. In response, system operators seek new sources of flexibility to maintain grid stability, with virtual power plants (VPPs) emerging as a promising solution~\cite{Riaz}. By aggregating diverse DERs, such as distributed generators (DG), heat pumps (HP), electric vehicles (EV), and battery energy storage systems (BESS), VPPs operate as unified entities that can interact with electricity markets similarly to conventional power plants.

Unlike conventional centralized generators, DERs are inherently uncertain, exhibiting volatility in weather-dependent generation, thermal demand, and user-driven behavior. These uncertainties affect both revenue streams and operational feasibility, particularly when physical network constraints are considered. Thus, it is critical to co-optimize market participation and DER dispatch under uncertainty. Moreover, unlocking DER flexibility requires the coordinated participation in sequential electricity markets (e.g., energy and reserve), which adds further layers of complexity for VPP operators.
The multi-market scheduling of VPPs under uncertainty has been addressed in the literature mainly through two modeling approaches: 1) stochastic optimization, and 2) robust optimization. 

Within the stochastic optimization stream,~\cite{Tajeddini2014} proposed a two-stage model for day-ahead and imbalance market participation, capturing uncertainty in renewable generation and market prices. This framework was subsequently extended to joint energy and reserve market scheduling, incorporating additional uncertainties such as reserve activation and balancing settlement~\cite{Dabbagh2016}. Later studies further expanded the market stack jointly considering energy, reserve capacity, and reserve activation decisions~\cite{Baringo2019,Vahedipour-Dahraie2021}. More recent work adopted multistage stochastic programs to reflect the sequential timing of electricity markets and the progressive revelation of uncertainty~\cite{Kraft2023,Fusco2023}. These problems are solved via the scenario approximation approach, where the underlying uncertainty is discretized into a finite set of scenarios, yielding a large-scale deterministic equivalent that is optimized. Such an approach leads to rapid growth in model size and memory as the number of scenarios increases. Accordingly, several studies address this through scenario reduction~\cite{Falabretti2023} and decomposition techniques~\cite{Zapparoli2026_VPP_scheduling,Fusco2023}, yet practical applications remain limited to small-scale VPPs.

Robust optimization (RO) was introduced in VPP scheduling primarily to improve computational tractability. RO characterizes uncertainty via deterministic sets and optimizes for the best performance under the worst-case realization, yielding a low-dimensional robust counterpart that can be solved efficiently. Early RO formulations established bidding tools for VPPs in day-ahead and real-time markets, using uncertainty sets to hedge renewable production and price forecast errors~\cite{Rahimiyan2016}. Building on this, subsequent work focused on improving scalability through decomposition-based solution schemes~\cite{Zheming2016}. Authors in~\cite{Liu2022} extend the robust formulation to include power reserve and carbon trading. 
A key limitation of RO is the tendency to produce overly conservative results. To mitigate this issue,~\cite{Ma2025} proposes a data-driven approach to construct sets that better align with empirical forecast errors. Building on this,~\cite{Nemati2025_2,Nemati2025_3} introduces flexible robust formulations that exploit uncertainty budgets and asymmetric set definitions to moderate worst-case protection across interacting uncertainties and sequential market stages. More recently,~\cite{Nemati2025_1} proposes a regret-based robust framework that explicitly quantifies the economic cost of conservatism, providing an interpretable parameter for conservatism tuning. However, these refinements only mitigate conservatism through ex-ante tuning and cannot capture the true probabilistic nature of the process. Consequently, RO remains structurally suboptimal.

Recent literature proposes polynomial chaos expansion (PCE) as an approach to preserve the stochastic structure while mitigating the scalability issues of scenario approximation. Analogous to how a Fourier series characterizes a signal through the amplitudes of its frequency components, PCE represents random variables through the amplitudes of orthogonal polynomials over the probability domain~\cite{Xiu2010}. In this way, uncertainty is described through polynomial coefficients rather than samples. By projecting the model's equations directly onto these polynomial functions, the intrusive variant of PCE transforms the stochastic problem into a deterministic spectral counterpart that solves for the amplitude of each polynomial term. This preserves the probabilistic structure while avoiding scenario explosion, yielding a lower-dimensional and more tractable solution.

Intrusive PCE has been established in the uncertainty quantification literature~\cite{Xiu2010} and later adopted in the power systems domain~\cite{Roald2023}. Authors in~\cite{Mhlpfordt2018} used intrusive PCE to construct a general framework for stochastic AC optimal power flow (OPF). The approach projects the full AC power flow equations in PCE space~\cite{Muhlpfordt2019}, showing excellent optimality and computational performance. To improve scalability, computational enhancements were proposed, such as the use of sparsity~\cite{Mtivier2020} and alternative network constraints formulations~\cite{VanAcker2022}. Recent works further extended intrusive PCE to solve the stochastic OPF in hybrid AC/DC grids under continuous non-Gaussian uncertainty~\cite{Yurtseven2025_OPF} and to the stochastic transmission switching problem~\cite{Yurtseven_TSP}. At the distribution level, intrusive PCE has been applied to compute stochastic PV hosting capacity~\cite{Koirala2024_PVHC}, day-ahead dynamic operating envelopes~\cite{Koirala2024_DOE}, and probabilistic locational marginal prices~\cite{Austnes2025} on realistic LV feeders.
Despite these advancements, existing power system applications of intrusive PCE primarily address physical grid operations. To the best of the authors' knowledge, this spectral approach has not been applied to market scheduling problems, which differ in scope and structure. Furthermore, the broader adoption of intrusive PCE is restricted by implementation complexity. Existing open-source packages either provide fundamental polynomial algebra~\cite{Mhlpfordt2020_polychaos} or implement the complete spectral reformulation just for specialized applications~\cite{StochasticPowerModelsjl}. Consequently, intrusive PCE remains difficult to reproduce, adapt, and deploy beyond specific use cases.

Overall, existing VPP scheduling frameworks exhibit a tradeoff between the computational intractability of scenario-based stochastic programs and the inherent conservatism of robust optimization. Intrusive PCE offers an alternative to preserve probabilistic optimality while enhancing computational scalability. However, a spectral reformulation of multi-market VPP scheduling has not yet been explored, and no dedicated coefficient-space reformulation has been proposed for this problem class. Finally, the absence of an application-agnostic projection tool for intrusive PCE limits its practical implementation.

To address these gaps, this paper proposes a novel stochastic multi-market scheduling framework. It preserves the economic optimality of stochastic optimization while improving computational tractability using intrusive PCE.
The proposed framework models the multi-market operation of a VPP participating in the day-ahead energy and reserve capacity markets, while accounting for real-time DER operations, network constraints, reserve activation, and imbalance settlement. The problem is formulated as a two-stage stochastic program that jointly optimizes market bids and DER dispatch under uncertainty. By projecting the resulting formulation into the PCE domain, an efficient deterministic coefficient-based counterpart is obtained. The approach is evaluated on a realistic Swiss low-voltage network with a comprehensive portfolio of flexible resources, including DGs, EVs, HPs, and BESS. Its computational performance and solution quality are benchmarked against a state-of-the-art scenario-based stochastic formulation solved via Benders decomposition.

The main contributions of this paper are threefold. First, we propose an intrusive-PCE reformulation of the multi-market VPP scheduling problem, in which the second-stage decisions are modeled in the spectral domain. The resulting spectral formulation preserves the economic optimality of stochastic programming while improving computational scalability, enabling VPP operators to manage large-scale DER portfolios with limited computational effort.
Second, we demonstrate the developed framework in a realistic case study and show that it achieves solution quality comparable to scenario-based stochastic optimization while significantly reducing computational time and resource requirements.
Third, we release an open-source, application-agnostic projection tool~\cite{SpectralStochOpt} that automates the derivation of deterministic spectral counterparts for generic single- and two-stage stochastic programs. By lowering the implementation barrier for intrusive PCE, the tool supports computational reproducibility and facilitates the adoption of spectral methods across a broad range of applications.

The remainder of this paper is organized as follows. Section~\ref{sec:method} describes the proposed method. Section~\ref{sec:cstudy} introduces the case study used to demonstrate the method. Section~\ref{sec:results} provides results followed by a discussion in Section~\ref{sec:discussion}. Section~\ref{sec:conclusions} concludes and provides an outlook for future work.

\section{Method}
\label{sec:method}
\begin{figure*}
    \centering
    \includegraphics[width=\textwidth]{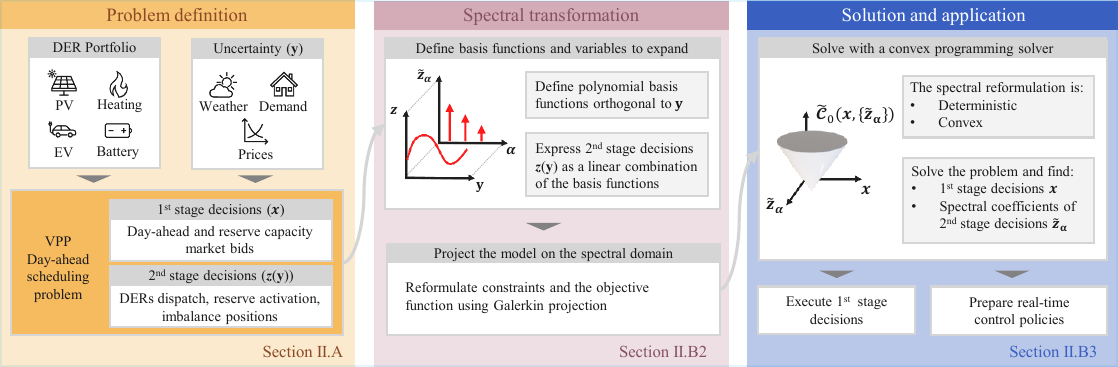}
    \caption{High-level description of the methodology for VPP multi-market scheduling under uncertainty.}
    \label{fig:methodology_overview}
\end{figure*}

This section details the framework for the VPP multi-market scheduling under uncertainty. The proposed approach leverages intrusive PCE to transform the stochastic decision process into a tractable, deterministic optimization problem in the spectral domain. Section~\ref{sec:method_scheduling_model} formulates the VPP scheduling model as a two-stage stochastic program. Section~\ref{sec:method_PCE} introduces the fundamental concepts of PCE and details the spectral reformulation of the objective function and constraints. Fig.~\ref{fig:methodology_overview} illustrates a high-level overview of the methodology.

\subsection{VPP Scheduling problem}
\label{sec:method_scheduling_model}
The VPP scheduling tool must coordinate a heterogeneous portfolio of DERs to act as a single entity in energy and reserve markets. The model is formulated as a two-stage stochastic program following the temporal sequence of market clearing utilized in European electricity markets. The operator decides, in the first stage, the quantities to bid in the day-ahead and the reserve capacity market. Simultaneously, the VPP operator accounts for the uncertainty in the next-day forecasts of prices and DER operations. In the second stage, which represents the delivery day, the operator decides how to operate the flexible DERs, how to bid in the reserve activation market, and which imbalance position to take.
We assume the VPP is a price taker in all markets, facing exogenous prices for energy, reserve capacity, and reserve activation.

Let $\boldsymbol{x} \in \mathbb{R}^{N^x}$ be the vector of first-stage decisions, representing day-ahead energy and reserve capacity offers. The continuous random vector $\mathbf{y} \in \mathbb{R}^{N^\omega}$ comprises uncertain parameters, including market-clearing prices, renewable energy availability, and load forecasts. After uncertainty is realized, the second-stage decisions $\boldsymbol{z} \in \mathbb{R}^{N^z}$ are taken and include reserve activations, imbalance quantities, and DER dispatch variables.
The VPP operator aims to define the optimal day-ahead bids $\boldsymbol{x}$ that perform the best against market and operational volatility. Under a risk-neutral perspective, this is achieved by minimizing the expected total operating cost $f(\boldsymbol{x}, \mathbf{y})$, which depends on both the bidding decisions and the realization of uncertain parameters: 
\begin{equation} 
\label{eq:general_formulation} 
\min_{\boldsymbol{x}} \  \mathbb{E} \left[ f(\boldsymbol{x}, \mathbf{y}) \right]. 
\end{equation} 
For a given realization of uncertainty $\mathbf{y}$, the total operating cost corresponds to the optimal value of a second-stage operational problem, in which the VPP adjusts its dispatch, reserve activation, and imbalance positions subject to financial commitments and physical feasibility:
\begin{equation}
\label{eq:second_stage_problem}
\begin{aligned}
f(\boldsymbol{x},\mathbf{y}) = \min_{\boldsymbol{z}} \quad & \mathcal{C}(\boldsymbol{x},\boldsymbol{z},\mathbf{y}) \\
\text{s.t.} \quad & \boldsymbol{g}(\boldsymbol{x},\boldsymbol{z},\mathbf{y})=\boldsymbol{0}\\
& \boldsymbol{h}(\boldsymbol{x},\boldsymbol{z},\mathbf{y})\le \boldsymbol{0},
\end{aligned}
\end{equation}
where $\mathcal{C}(\cdot)$ represents the total operating cost, while $\boldsymbol{g}(\cdot)$ and $\boldsymbol{h}(\cdot)$ denote the sets of equality and inequality constraints. The specific mathematical formulations are consistent with~\cite{Zapparoli2026_VPP_scheduling}. 

The operating cost $\mathcal{C}(\cdot)$ aggregates DERs' operational costs, network tariffs, and the revenues generated from the sequential energy and reserve markets. The constraint sets enforce both market rules and physical feasibility. From a market perspective, they: 1) couple the multi-market schedule to the VPP's actual physical power exchange via the imbalance position; 2) limit reserve activation bids to prequalified capacities; and 3) enforce that the contracted reserve capacity is offered in the reserve activation market. Physically, they guarantee the operational limits of the DERs in the portfolio and the distribution grid. At the device level, the linear constraints model: 1) the state-of-charge dynamics, power limits, and conversion efficiencies for BESS and EVs; 2) the thermal comfort boundaries and active power limits for HPs; and 3) the active and reactive power limits for DGs. At the network level, the linear DistFlow equations enforce nodal power balance, bus voltage limits, and branch thermal capacity constraints. Since the objective function, market constraints, and physical constraints are linear, the second-stage problem is a linear program for any given realization of the uncertainty. A detailed formulation of the objective and constraints is provided in to~\cite{Zapparoli2026_VPP_scheduling}.

The formulation in~\eqref{eq:general_formulation}-\eqref{eq:second_stage_problem} is a two-stage stochastic program with continuous random variables. This problem cannot be solved directly, because the continuous support of $\mathbf{y}$ implies an infinite-dimensional uncertainty space, requiring the evaluation of an infinite number of second-stage decisions and constraints. To make the problem solvable, uncertainty must be discretized.

The conventional approach is scenario approximation, where the probability distribution of $\mathbf{y}$ is approximated using a finite set of $N^s$ discrete scenarios. 
The second-stage problem is repeated for every scenario, yielding a large-scale deterministic program. The dimensionality explodes as the number of scenarios increases, making it difficult to adequately capture the uncertainty. To improve tractability, decomposition techniques are typically employed, most notably Benders decomposition, allowing for parallel computing~\cite{Zapparoli2026_VPP_scheduling,Fusco2023}.

In this paper, we propose an alternative approach to discretize the uncertainty space. Rather than assuming that the model behavior over the uncertainty domain can be accurately represented through a finite number of discrete realizations, we assume that it can be captured by a finite sum of polynomial functions. This is still a discretization, since a continuous uncertain parameter or decision variable is now entirely defined by a finite set of deterministic coefficients associated with the defined polynomial basis functions. The details on PCE are provided in the following section.

\subsection{Polynomial chaos expansion}
\label{sec:method_PCE}
Polynomial chaos expansion provides a spectral representation of the system's behavior over the uncertainty domain. Rather than evaluating the model at a finite set of sampled realizations, PCE represents stochastic variables as a linear combination of polynomial modes defined over the probability space of the uncertain parameters~\cite{Xiu2010}. Conceptually, this is analogous to a Fourier series expansion. While Fourier analysis decomposes a signal into orthogonal trigonometric modes in the time domain, PCE decomposes the system response into orthogonal polynomial modes in the uncertainty domain. For a large class of models, exhibiting smooth dependence on uncertain inputs, low-order polynomial expansions are often sufficient to accurately approximate the stochastic response. As a result, PCE enables an efficient representation of uncertainty, reducing dimensionality and computational complexity~\cite{Muhlpfordt2019}.

PCE is generally implemented in two variants: non-intrusive and intrusive. Non-intrusive techniques estimate spectral coefficients by treating the model as a functional black box, typically utilizing sampling, regression, or stochastic collocation~\cite{Xiu2010}. In contrast, this study employs the intrusive variant, where the governing equations of the VPP scheduling model are directly projected into the PCE domain using the stochastic Galerkin method. This transformation maps the original stochastic program into a deterministic system of coupled equations, with the PCE coefficients as decision variables. The spectral problem can be solved by propagating the uncertainty and finding the optimal solution at once.

The remainder of this section is organized as follows: first, Section~\ref{sec:PCE_basis} defines the orthogonal polynomial modes; next, Section~\ref{sec:pce_reformulation} details the spectral reformulation of the two-stage stochastic program; and finally, Section~\ref{sec:pce_properties} discusses the properties of the resulting deterministic counterpart.

\subsubsection{Definition of the polynomial modes}
\label{sec:PCE_basis}
Let $\mathbf{y} \in \mathbb{R}^{N^\omega}$ denote the vector of the uncertain parameters introduced in Section~\ref{sec:method_scheduling_model}. To apply PCE, the parameters must be uncorrelated; if correlations exist, copula-based transformations can be applied to map $\mathbf{y}$ to a set of independent random variables~\cite{MARA2021107795}. Under this assumption, $\mathbf{y}$ is characterized by a joint probability density function (PDF) $\rho(\mathbf{y})=\prod_{i=1}^{N^\omega}\rho_i(\mathrm{y}_i)$. For each independent component $\mathrm{y}_i$, a univariate family of orthogonal polynomials $\{\psi_{n}^{(i)}(\mathrm{y}_i)\}_{n \in \mathbb{N}}$ is selected, where $n$ indicates the degree of the polynomial. These polynomials are defined to be orthogonal with respect to the marginal PDF $\rho_i(\mathrm{y}_i)$.
\begin{equation}
\label{eq:orthogonality}
\begin{aligned}
\langle \psi_{n}^{(i)}(\mathrm{y}_i), \psi_{m}^{(i)}(\mathrm{y}_i) \rangle 
& = \int_{\Omega_i} \psi_{n}^{(i)}(\mathrm{y}_i)\psi_{m}^{(i)}(\mathrm{y}_i)\rho_i(\mathrm{y}_i)\,d\mathrm{y}_i\\
& = \gamma_{n}^{(i)} \delta_{nm},
\end{aligned}
\end{equation}
where $\Omega_i$ is the support of $\mathrm{y}_i$, $\langle \cdot,\cdot \rangle$ denotes the inner product,  $\gamma_{n}^{(i)}$ is the norm of the polynomial $\psi_{n}^{(i)}(\mathrm{y}_i)$, and $\delta_{nm}$ is the Kronecker delta, which equals 1 if $n = m$ and 0 otherwise. When an orthonormal basis is adopted, $\gamma_{n}^{(i)}=1$. Practically, for standard distributions, the polynomial bases are selected according to the Askey scheme~\cite{Sullivan2015UQ}. Examples are Hermite polynomials for Gaussian distributions, Legendre polynomials for uniform distributions, Laguerre polynomials for Gamma distributions, and Jacobi polynomials for Beta distributions.

Once the univariate bases for each uncertain parameter $\mathrm{y}_i$ are determined, the multivariate polynomial basis functions are constructed via tensor products of the univariate bases. Let $\boldsymbol{\alpha} = (\alpha_1, \dots, \alpha_{N^\omega}) \in \mathbb{N}^{N^\omega}$ be a multi-index, where each component $\alpha_i$ represents the degree of the univariate polynomial basis of the $i$-th germ. The corresponding multivariate basis function $\Psi_{\boldsymbol{\alpha}}(\mathbf{y})$ is defined as:
\begin{equation}
\label{eq:multivariate_basis}
\Psi_{\boldsymbol{\alpha}}(\mathbf{y}) = \prod_{i=1}^{N^\omega} \psi_{\alpha_i}^{(i)}(\mathrm{y}_i).
\end{equation}

Any finite-variance random variable, such as a second-stage decision variable $z(\mathbf{y})$, can be approximated by a truncated polynomial chaos expansion:
\begin{equation}
\label{eq:pce_expansion}
z(\mathbf{y}) \approx \sum_{\boldsymbol{\alpha} \in \mathcal{A}} \tilde{z}_{\boldsymbol{\alpha}} \Psi_{\boldsymbol{\alpha}}(\mathbf{y}),
\end{equation}
where $\tilde{z}_{\boldsymbol{\alpha}}$ are deterministic coefficients and $\mathcal{A} \subset \mathbb{N}^{N^\omega}$ is the set of considered multi-indices. For a total-degree truncation scheme, this set is defined as:
\begin{equation}
\label{eq:multi_index_set}
\mathcal{A} = \left\{ \boldsymbol{\alpha} \in \mathbb{N}^{N^\omega} : \sum_{i=1}^{N^\omega} \alpha_i \le N^d \right\},
\end{equation}
where $N^d$ is the maximum total degree of the expansion. The cardinality of the basis, which represents the total number of bases, is given by:
\begin{equation}
\label{eq:pce_cardinality}
|\mathcal{A}| = \binom{N^\omega + N^d}{N^d} = \frac{(N^\omega + N^d)!}{N^\omega! \, N^d!}.
\end{equation}
The approximation converges as $N^d$ increases.

We are expressing $z(\mathbf{y})$ as a polynomial function of the uncertain parameters. In principle, any polynomial basis could be used; however, the choice of basis functions orthonormal to the input PDFs allows us to express the statistical moments of the expanded variable just as a function of the coefficients $\tilde{z}_{\boldsymbol{\alpha}}$~\cite{Xiu2010}. In particular, the expected value and variance of $z(\mathbf{y})$ are given by:
\begin{equation}
\label{eq:expected_value}
\mathbb{E}[z(\mathbf{y})] = \tilde{z}_{\mathbf{0}},
\end{equation}
\begin{equation}
\label{eq:variance}
\mathrm{Var}[z(\mathbf{y})] = \sum_{\boldsymbol{\alpha} \in \mathcal{A} \setminus \{\mathbf{0}\}} \tilde{z}_{\boldsymbol{\alpha}}^2,
\end{equation}
where $\mathbf{0} = (0, \dots, 0)$ is the multi-index of the constant basis function.

\subsubsection{Two-stage problem reformulation}
\label{sec:pce_reformulation}
To reformulate the two-stage stochastic program in \eqref{eq:general_formulation}--\eqref{eq:second_stage_problem} via PCE, the second-stage (wait-and-see) variables are expanded onto the polynomial basis, whereas the first-stage (here-and-now) decisions are kept deterministic. The equality constraints in \eqref{eq:second_stage_problem} are preserved in the projected model, whereas the inequality constraints are treated through a chance-constraint approximation. Together, these reformulation steps retain the structure of the original problem and yield a fully deterministic optimization problem in the coefficient space. This problem can then be solved using standard mathematical programming techniques.

\textit{Decision variables.}
Because the first-stage decisions $\boldsymbol{x}\in\mathbb{R}^{N^x}$ are taken prior to the realization of uncertainty, they are independent of $\mathbf{y}$. Therefore, in the spectral domain, they are assigned the constant mode exclusively:
\begin{equation}
\label{eq:x_constant_mode}
\boldsymbol{x}=\tilde{\boldsymbol{x}}_{\mathbf{0}},
\qquad
\tilde{\boldsymbol{x}}_{\boldsymbol{\alpha}}=\boldsymbol{0}\;\;\forall \boldsymbol{\alpha}\in\mathcal{A}\setminus\{\mathbf{0}\}.
\end{equation}
Conversely, the second-stage decisions $\boldsymbol{z}\in\mathbb{R}^{N^z}$ adapt to the realized uncertainty and are therefore approximated using the truncated PCE in~\eqref{eq:pce_expansion}. The associated coefficient vectors are denoted by $\{\tilde{\boldsymbol{z}}_{\boldsymbol{\alpha}}\}_{\boldsymbol{\alpha}\in\mathcal{A}}$.

\textit{Equality constraints.}
Consider an equality constraint $g_j(\boldsymbol{x},\boldsymbol{z}(\mathbf{y}),\mathbf{y})=0$ from $\boldsymbol{g}(\cdot)=\boldsymbol{0}$ in \eqref{eq:second_stage_problem}. This equality is enforced in the PCE domain via the stochastic Galerkin method, which requires the constraint residual to be orthogonal to each basis function, i.e.,
\begin{equation}
\label{eq:galerkin_proj}
\tilde{g}_{j_{\boldsymbol{\alpha}}}(\boldsymbol{x}, \{\tilde{\boldsymbol{z}}_{\boldsymbol{\alpha}}\}) = \left\langle g_j(\boldsymbol{x},\boldsymbol{z}(\mathbf{y}),\mathbf{y}), \Psi_{\boldsymbol{\alpha}} \right\rangle = 0,
\qquad \forall \boldsymbol{\alpha}\in\mathcal{A}
\end{equation}
where $\tilde{g}_{j_{\boldsymbol{\alpha}}}(\boldsymbol{x}, \{\tilde{\boldsymbol{z}}_{\boldsymbol{\alpha}}\})$ is the residual of the projected equality constraint for the multi-index $\boldsymbol{\alpha}$. It is an analytical function of the PCE coefficients and the first-stage decisions. Therefore, each equality constraint in the second stage yields $|\mathcal{A}|$ deterministic equality constraints in the PCE domain.

For linear relations, the Galerkin projection reduces to coefficient-wise identities. For example, let $u(\mathbf{y})$ and $v(\mathbf{y})$ denote two second-stage variables. Then, for the linear constraint $u(\mathbf{y}) + v(\mathbf{y}) = 0$ the projection yields $\tilde{u}_{\boldsymbol{\alpha}} + \tilde{v}_{\boldsymbol{\alpha}} = 0$ for all $\boldsymbol{\alpha}\in\mathcal{A}$. For bilinear constraints, e.g., $u(\mathbf{y})v(\mathbf{y}) = 0$, the projection is performed through the triple-product tensor $M_{\boldsymbol{\zeta},\boldsymbol{\eta},\boldsymbol{\alpha}}$, as:
\begin{equation}
\label{eq:mult_tensor}
\sum_{\boldsymbol{\zeta}\in\mathcal{A}}\sum_{\boldsymbol{\eta}\in\mathcal{A}}
M_{\boldsymbol{\zeta},\boldsymbol{\eta},\boldsymbol{\alpha}}\,(\tilde{u}_{\boldsymbol{\zeta}} \tilde{v}_{\boldsymbol{\eta}}) = 0
\end{equation}
where:
\begin{equation}
\label{eq:triple_product_tensor_definition}
M_{\boldsymbol{\zeta},\boldsymbol{\eta},\boldsymbol{\alpha}}=\langle \Psi_{\boldsymbol{\zeta}}\Psi_{\boldsymbol{\eta}},\Psi_{\boldsymbol{\alpha}}\rangle.
\end{equation}
The tensor depends only on the chosen basis and can be precomputed offline. This procedure preserves the algebraic structure of the original constraints.

\textit{Inequality constraints.}
Consider an inequality constraint $h_j(\boldsymbol{x},\boldsymbol{z}(\mathbf{y}),\mathbf{y})\leq0$ from $\boldsymbol{h}(\cdot)\leq\boldsymbol{0}$ in \eqref{eq:second_stage_problem}. To obtain a tractable reformulation in the coefficient domain, we approximate each inequality constraint by a chance constraint with violation probability $\epsilon\in(0,1)$:
\begin{equation}
\label{eq:chance_constraint}
\mathbb{P}\!\left(h_j(\boldsymbol{x},\boldsymbol{z}(\mathbf{y}),\mathbf{y})\leq0\right)\ge 1-\epsilon.
\end{equation}
To solve the chance constraint, we adopt the moment-based approximation:
\begin{equation}
\label{eq:soc_chance}
\mathbb{E}[h_j(\boldsymbol{x},\boldsymbol{z}(\mathbf{y}),\mathbf{y})] + \lambda(\epsilon) \sqrt{\mathrm{Var}[h_j(\boldsymbol{x},\boldsymbol{z}(\mathbf{y}),\mathbf{y})]} \le 0,
\end{equation}
where $\lambda(\epsilon)$ is a safety factor chosen based on a distributional assumption or a conservative bound. 
To obtain a tractable deterministic counterpart, we project $h_j(\boldsymbol{x},\boldsymbol{z}(\mathbf{y}),\mathbf{y})$ onto the PCE basis as in \eqref{eq:galerkin_proj}, yielding the coefficients expression $\tilde{h}_{j_{\boldsymbol{\alpha}}}(\boldsymbol{x}, \{\tilde{\boldsymbol{z}}_{\boldsymbol{\alpha}}\})$. Recalling the properties from \eqref{eq:expected_value}--\eqref{eq:variance}, we reformulate the chance constraint as a function of the PCE coefficients, producing a second-order cone (SOC) constraint:
\begin{equation}
\label{eq:soc_chance_coeff}
\tilde{h}_{j_{\mathbf{0}}}(\boldsymbol{x}, \{\tilde{\boldsymbol{z}}_{\boldsymbol{\alpha}}\}) + \lambda(\epsilon)\sqrt{\sum_{\boldsymbol{\alpha}\ne \mathbf{0}} \tilde{h}_{j_{\boldsymbol{\alpha}}}^2(\boldsymbol{x}, \{\tilde{\boldsymbol{z}}_{\boldsymbol{\alpha}}\})} \le 0.
\end{equation}
Importantly, \eqref{eq:soc_chance_coeff} maps each inequality constraint in the recourse into a single SOC constraint in the projected problem.

\textit{Objective function.}
The VPP operator minimizes the expected optimal second-stage cost, where the second stage is given by \eqref{eq:second_stage_problem}. The second stage cost is also a wait-and-see variable; therefore, it is expanded as in \eqref{eq:pce_expansion}:
\begin{equation}
\label{eq:C_pce}
\mathcal{C}\!\left(\boldsymbol{x},\boldsymbol{z}(\mathbf{y}),\mathbf{y}\right)
\approx \sum_{\boldsymbol{\alpha}\in\mathcal{A}}
\tilde{\mathcal{C}}_{\boldsymbol{\alpha}}\!\left(\boldsymbol{x},\{\tilde{\boldsymbol{z}}_{\boldsymbol{\alpha}}\}\right)\,
\Psi_{\boldsymbol{\alpha}}(\mathbf{y}),
\end{equation}
where the expression of the coefficients $\tilde{\mathcal{C}}_{\boldsymbol{\alpha}}(\cdot)$ is obtained by applying Galerkin projection to the objective function as in~\ref{eq:galerkin_proj}.
By taking the expectation of $\mathcal{C}\!\left(\boldsymbol{x},\boldsymbol{z}(\mathbf{y}),\mathbf{y}\right)$ with respect to the uncertain parameters and exploiting the orthonormality of the basis, the expected operating cost is given by the zero-order coefficient:
\begin{equation}
\label{eq:obj_pce}
\mathbb{E}\!\left[\mathcal{C}\!\left(\boldsymbol{x},\boldsymbol{z}(\mathbf{y}),\mathbf{y}\right)\right]
\approx
\tilde{\mathcal{C}}_{\mathbf{0}}\!\left(\boldsymbol{x},\{\tilde{\boldsymbol{z}}_{\boldsymbol{\alpha}}\}\right).
\end{equation}
Therefore, the projected risk-neutral objective is simply to minimize $\tilde{\mathcal{C}}_{\mathbf{0}}$, which represents the mean operating cost implied by the PCE-parametrized recourse.

The reformulation steps in \eqref{eq:orthogonality}--\eqref{eq:obj_pce} are automated by the application-agnostic projection tool \textit{SpectralStochOpt}~\cite{SpectralStochOpt}, released with this work. Given a Pyomo formulation of the original stochastic program, the tool i) generates the orthogonal polynomial basis, ii) constructs the corresponding coefficient-space representation of the recourse variables, iii) applies the Galerkin projection to the objective function and equality constraints, and iv) reformulates the inequality constraints through the moment-based SOC approximation. This yields the deterministic spectral counterpart of the original problem, directly solvable by standard optimizers.

\subsubsection{Properties of the projected model}
\label{sec:pce_properties}
Collecting \eqref{eq:x_constant_mode}--\eqref{eq:obj_pce}, the intrusive PCE reformulation yields the following deterministic coefficient-space approximation:
\begin{equation}
\label{eq:pce_socp_clean}
\begin{aligned}
\min_{\boldsymbol{x},\{\tilde{\boldsymbol{z}}_{\boldsymbol{\alpha}}\}_{\boldsymbol{\alpha}\in\mathcal{A}}}\quad
& \tilde{\mathcal{C}}_{\mathbf{0}}\!\left(\boldsymbol{x},\{\tilde{\boldsymbol{z}}_{\boldsymbol{\alpha}}\}\right) \\[1mm]
\text{s.t.}\quad
& \tilde{g}_{j_{\boldsymbol{\alpha}}}\!\left(\boldsymbol{x},\{\tilde{\boldsymbol{z}}_{\boldsymbol{\alpha}}\}\right)=0,
\hspace{1.3cm} \forall j\in\mathcal{J}_g,\ \forall \boldsymbol{\alpha}\in\mathcal{A},\\
& \tilde{h}_{j_{\mathbf{0}}}(\boldsymbol{x}, \{\tilde{\boldsymbol{z}}_{\boldsymbol{\alpha}}\})
+ \lambda(\epsilon)\sqrt{\sum_{\boldsymbol{\alpha}\ne \mathbf{0}}
\tilde{h}_{j_{\boldsymbol{\alpha}}}^2(\boldsymbol{x}, \{\tilde{\boldsymbol{z}}_{\boldsymbol{\alpha}}\})}
\le 0,\\[1mm]
& \hspace{5.6cm}
\forall j\in\mathcal{J}_h.
\end{aligned}
\end{equation}
where $\mathcal{J}_g$ and $\mathcal{J}_h$ are the index sets of the equality and inequality constraints in \eqref{eq:second_stage_problem}, respectively.

The decision variables of the projected model are the first-stage vector $\boldsymbol{x}$ and the PCE coefficients $\{\tilde{\boldsymbol{z}}_{\boldsymbol{\alpha}}\}_{\boldsymbol{\alpha}\in\mathcal{A}}$ associated with the second-stage decisions. Accordingly, each original recourse variable is replaced by $|\mathcal{A}|$ deterministic coefficients, each equality constraint generates $|\mathcal{A}|$ deterministic equalities through \eqref{eq:galerkin_proj}, and each inequality constraint is mapped to a single SOC constraint through \eqref{eq:soc_chance_coeff}. Since the underlying VPP scheduling model is linear and the inequality constraints are enforced through SOC reformulations, the resulting deterministic counterpart is a convex second-order cone program (SOCP).

For linear programs with linear objective and constraints, the optimal recourse policy is affine in the uncertainty~\cite{Mhlpfordt2018}. Hence, for the linear VPP operating model considered here, a first-degree expansion ($N^d=1$) is sufficient, yielding a cardinality $|\mathcal{A}| = N^\omega + 1$. This provides a compact representation whose size scales with the number of uncertainty sources.
Once the deterministic SOCP is solved, the optimal first-stage decisions and the second-stage coefficients are obtained simultaneously. The first-stage decisions can be used as-is to place the bids on the DAM and RCM, while the second-stage actions corresponding to a realization $\mathbf{y}=\boldsymbol{\omega}$ are obtained by evaluating the polynomial mapping:
\begin{equation}
\label{eq:z_eval}
\boldsymbol{z}(\boldsymbol{\omega}) \approx \sum_{\boldsymbol{\alpha}\in\mathcal{A}} \tilde{\boldsymbol{z}}_{\boldsymbol{\alpha}}\,\Psi_{\boldsymbol{\alpha}}(\boldsymbol{\omega}),
\end{equation}
which is computationally negligible and avoids re-optimization in real-time operations.

\section{Case study}
\label{sec:cstudy}
The following provides an overview of the considered case study, the uncertainty modeling, and the computational aspects of the implementation.

\subsection{Case study description}
\label{sec:cstudy_description}
\begin{figure}
    \centering
    \includegraphics[width=\linewidth]{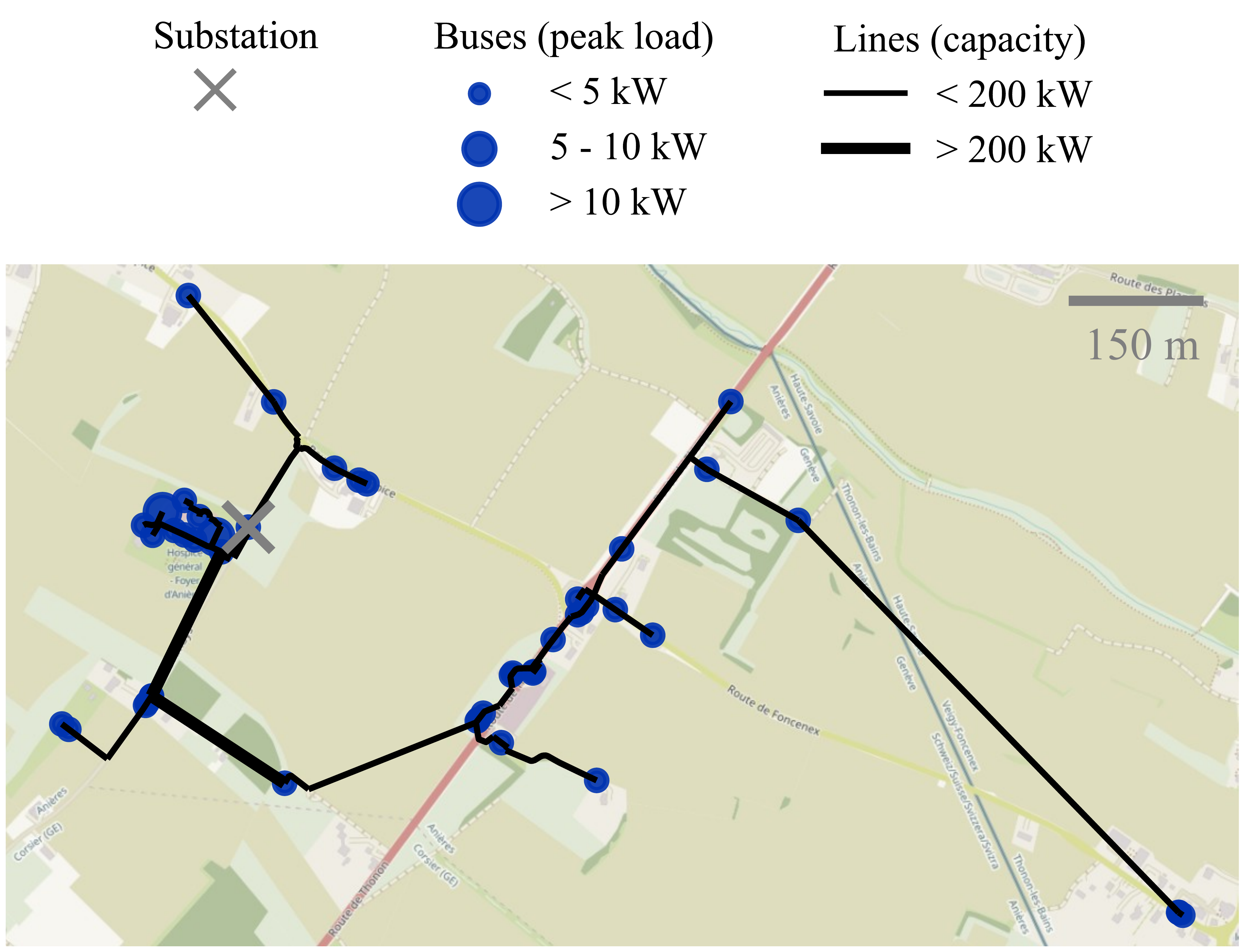}
    \caption{Case study network map showing the 77 buses, lines, and substation location.}
    \label{fig:network_map}
\end{figure}

The proposed framework is tested on a DER-based VPP operating on a low-voltage distribution network in Switzerland. The network topology is taken from the open-source Swiss distribution-grid dataset in~\cite{Oneto_2023}, while the nodal allocation of DER technologies and their operating conditions are adopted from~\cite{Zapparoli2025}. In this study, we consider the grid 6602-1\_0\_3~\cite{Oneto_2023}, shown in Fig.~\ref{fig:network_map}. It comprises 77 buses and represents a realistic residential-dominated LV system.
The VPP includes \SI{83}{kW} of peak non-dispatchable demand, \SI{375}{kWp} of rooftop PV capacity, \SI{147}{kW} of HP capacity, \SI{305}{kW} of battery storage with an average energy-to-power ratio of \SI{2.5}{h}, and daily EV demand of \SI{376}{kWh}. All DERs provide flexibility subject to their operational constraints.

We consider a day-ahead scheduling problem in which the VPP determines market offers for the following day over a \SI{24}{h} horizon under forecast uncertainty. Real-time operation is modeled with hourly resolution, which is used consistently for DER dispatch, reserve activation-market decisions, and imbalance settlement. Reserve capacity products are modeled with a \SI{4}{h} delivery duration. In this study, the VPP participates in the capacity and activation markets only for secondary reserve (aFRR).

For the market data, we adopted publicly available Swiss day-ahead prices~\cite{entsoe_transparency} and Swiss ancillary-service activation prices~\cite{swissgrid_tenders}. Since Swiss reserve capacity prices are not publicly accessible, we use German reserve capacity prices as a proxy~\cite{regelleistung2025}. Network and system tariffs are set to $c_t^{\text{tariff}}=\SI{206.5}{CHF/MWh}$ following~\cite{ewz_2024_tariff}. For all historical data, 2024 values are used. We select March 13th as the simulation day. This day reflects typical spring or autumn conditions in which all DERs are active.

\subsection{Uncertainty modeling}
\label{sec:cstudy_uncertainty_modeling}
Key non-financial inputs subject to forecasting uncertainty include: ambient temperature, solar irradiance, baseline non-dispatchable load profiles, and EV demand. Given the spatial proximity of the assets and the short-term scheduling horizon, we adopt the simplifying assumption that each forecasting error type affects all corresponding devices uniformly and consistently across all time steps. Under this assumption, non-financial uncertainty is represented by four lumped error variables: $\xi^{\mathrm{G}}$ for solar generation, $\xi^{\mathrm{T}}$ for ambient temperature, $\xi^{\mathrm{EV}}$ for EV demand, and $\xi^{\mathrm{L}}$ for non-dispatchable demand~\cite{zapparoli2025powerreservecapacityvirtual}.

For the market inputs, we model forecasting errors of day-ahead market and ancillary services prices. We assume these errors are statistically independent of the non-financial uncertainties and mutually independent. This is resulting in the lumped variables $\xi^{\mathrm{DAM}}$ for the day-ahead market prices, $\xi^{\mathrm{RCM,aFRR}}$ for both upward and downward secondary reserve capacity prices, and $\xi^{\mathrm{RAM,aFRR,up}}$ and $\xi^{\mathrm{RAM,aFRR,dn}}$ for the secondary reserve upward and downward activation prices, respectively.
The detailed mapping of the lumped uncertain parameters to the model inputs is described in~\cite{Zapparoli2026_VPP_scheduling}. All uncertainty parameters are calibrated through literature-based standard deviations, summarized in Table~\ref{tab:Forecasting errors}.

\begin{table}
\caption[Parameters for forecasting errors]{Uncertainty model parameters for the considered forecasting errors.}
\label{tab:Forecasting errors}
\vspace{6pt}
\centering
\begin{tabular}{|c|c|c|c|c|} 
\hline
\textbf{Error} & \textbf{PDF} & \textbf{Standard deviation} & \textbf{Mean} & \textbf{Ref.} \\
\hline
$\xi^{\mathrm{L}}$ & Normal & 10.75\% & 0 &~\cite{Kong} \\  
$\xi^{\mathrm{G}}$ & Normal & 8.15\% & 0 &~\cite{zapparoli2025powerreservecapacityvirtual}\\  
$\xi^{\mathrm{T}}$ & Normal & 1.5\,K & 0 &~\cite{zapparoli2025powerreservecapacityvirtual}\\    
$\xi^{\mathrm{EV}}$ & Uniform & 5.77\% & 10\% &~\cite{zapparoli2025powerreservecapacityvirtual}\\ 
$\xi^{\mathrm{DAM}}$ & Normal & 4.28\,EUR/MWh & 0 &~\cite{Lago} \\  
$\xi^{\mathrm{RCM,aFRR}}$ & Normal & 3.30\,EUR/MW & 0 &~\cite{Cardo-Miota2023}\\  
$\xi^{\mathrm{RAM,aFRR,up}}$ & Normal & 32.08\,EUR/MWh & 0 &~\cite{Failing2025} \\ 
$\xi^{\mathrm{RAM,aFRR,dn}}$ & Normal & 21.25\,EUR/MWh & 0 &~\cite{Failing2025} \\ 
\hline
\end{tabular}
\end{table}

\subsection{Computational setup}
\label{sec:cstudy_computational_aspects}
For the intrusive PCE reformulation, we adopt a total-degree truncation of order $N^d=1$. Since the uncertainty vector has $N^\omega=8$ independent components, the resulting basis cardinality is $|\mathcal{A}|=N^\omega+1=9$.
Inequality constraints are enforced through the moment-based SOC approximation in \eqref{eq:soc_chance_coeff} with safety factor $\lambda=1.645$, which corresponds to a violation probability $\epsilon=1-\Phi(1.645)\approx 5\times 10^{-2}$ under a Gaussian residual assumption~\cite{Yurtseven_TSP}. All the reformulation is performed via the released projection tool \textit{SpectralStochOpt}~\cite{SpectralStochOpt}, which maps the original model into the coefficient-space SOCP prior to solution.

The proposed intrusive PCE formulation is benchmarked against a state-of-the-art scenario approximation using Latin hypercube sampling and Benders decomposition~\cite{Zapparoli2026_VPP_scheduling}. For the scenario approximation, we generate $N^s\in\{50,\dots,1000\}$ second-stage scenarios from the uncertainty model in Table~\ref{tab:Forecasting errors}. For each $N^s$, the experiment is repeated 20 times with different realizations to quantify sampling variability. A separate set of runs with $N^s=2000$ (20 repetitions) is performed to construct a high-accuracy reference: the resulting objective estimates exhibit a coefficient of variation of \SI{0.2}{\%} across repetitions, and the corresponding 20-run average is used as the reference for accuracy evaluation.

All simulations are executed on the ETH Zurich \textit{Euler} high-performance computing cluster. The scenario-based benchmark is solved using Benders decomposition with parallelization via Message Passing Interface (MPI) across 50 single-core processes, whereas the intrusive PCE formulation is solved as a single monolithic SOCP on a single core. Both formulations use Gurobi as a solver. 

\section{Results}
\label{sec:results}
\begin{figure*}
    \centering
    \includegraphics[width=\textwidth]{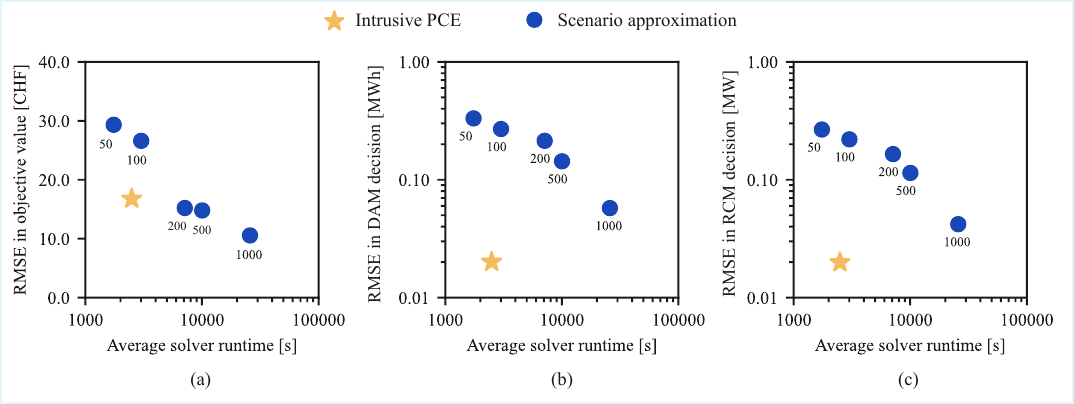}
    \caption{Accuracy--runtime comparison between intrusive PCE and scenario approximation (SA). RMSE with respect to the 2000 samples SA reference is shown for (a) objective value, (b) day-ahead energy (DAM) offers, and (c) reserve-capacity (RCM) offers. SA markers are labeled by the number of scenarios.}
    \label{fig:runtime_scatters}
\end{figure*}

This section benchmarks the proposed intrusive PCE reformulation against the state-of-the-art scenario approximation (SA) solved via Benders decomposition. First, computational performance is compared. Next, the accuracy of the objective value and first-stage scheduling decisions is assessed against the SA reference. Finally, an empirical validation of the chance-constraint approximation is performed.

\subsection{Computational performance}
\label{sec:results_compute}
We compare the computational performance of the proposed intrusive PCE approach with SA in terms of wall-clock time, core-hours, and memory usage. Here, wall-clock time denotes the solution time, whereas core-hours quantify the total computational effort as the product of solution time and the number of allocated cores. The results are summarized in Table~\ref{tab:compute_summary}.

The intrusive PCE formulation is solved in two steps. First, the original model in the decision-variable domain is projected onto the spectral domain, which requires approximately \SI{1.5}{min}. This yields a deterministic SOCP with approximately $5.3\times10^5$ variables and constraints. The resulting SOCP is then solved in \SI{41.6}{min} on a single core, using approximately \SI{7}{GB} of RAM. The total runtime is therefore \SI{43.1}{min}, corresponding to 0.72~core-hours.

In contrast, SA exhibits wall-clock times comparable to intrusive PCE only at the smallest scenario sets ($N^s=$50-100), and the solution time increases sharply as the number of scenarios grows, reaching \SI{7.18}{h} at $N^s=1000$ and \SI{9.10}{h} for the $N^s=2000$ reference. These runtimes are obtained by parallelizing the Benders subproblems across 50 processes, whereas the intrusive PCE formulation is solved on a single core. The corresponding core-hour requirement of SA is therefore substantially larger, ranging from 24.25~core-hours at $N^s=50$ to 454.83~core-hours at $N^s=2000$, compared with only 0.72~core-hours for intrusive PCE. The memory usage is also substantially larger, ranging from \SI{39.7}{GB} at $N^s=50$ to \SI{1.55}{TB} at $N^s=2000$.

Overall, the intrusive PCE formulation is already competitive in wall-clock time with the smallest SA instances and becomes faster than SA for $N^s\geq100$, despite being solved on a single core. More importantly, it requires one to three orders of magnitude fewer core-hours across all tested SA configurations, while also using substantially less memory. These results highlight the computational efficiency of the proposed spectral reformulation.

\begin{table}[t]
\centering
\caption{Computational summary of intrusive PCE and scenario approximation (SA) with Benders decomposition. SA runtimes are averages over 20 repetitions on 50 processes.}
\label{tab:compute_summary}
\vspace{4pt}
\begin{tabular}{|l|c|c|c|c|c|}
\hline
\textbf{Method} & $\boldsymbol{N^s}$ & \textbf{Cores} & \textbf{\makecell{Wall-clock\\time \SI{}{[h]}}} & \textbf{\makecell{Core-hours\\\SI{}{[h]}}} & \textbf{\makecell{RAM\\\SI{}{[GB]}}} \\
\hline
PCE & --    & 1  & \SI{0.72}{} & \SI{0.72}{}   & \SI{7}{} \\
SA  & 50    & 50 & \SI{0.49}{} & \SI{24.25}{}  & \SI{39.7}{} \\
SA  & 100   & 50 & \SI{0.84}{} & \SI{41.75}{}  & \SI{82.4}{} \\
SA  & 200   & 50 & \SI{1.98}{} & \SI{98.92}{}  & \SI{157.3}{} \\
SA  & 500   & 50 & \SI{2.79}{} & \SI{139.42}{} & \SI{387.5}{} \\
SA  & 1000  & 50 & \SI{7.18}{} & \SI{358.75}{} & \SI{779.6}{} \\
SA  & 2000  & 50 & \SI{9.10}{} & \SI{454.83}{} & \SI{1550}{} \\
\hline
\end{tabular}
\end{table}

\subsection{Accuracy in objective value and first-stage decisions}
\label{sec:results_accuracy}
Beyond computational performance, we evaluate the accuracy of the intrusive PCE solution in terms of (i) the objective value and (ii) the first-stage decisions, using the $N^s=2000$ SA estimates as the benchmark.
Fig.~\ref{fig:runtime_scatters}a reports the trade-off between solution time and objective-value error. Intrusive PCE yields an objective value of \SI{611.2}{CHF}, while the $N^s=2000$ SA reference is \SI{594.4}{CHF}, resulting in a deviation of \SI{16.8}{CHF}, corresponding to approximately \SI{2.8}{\%}. This level of accuracy is comparable to that of SA at intermediate sample sizes, e.g., around $N^s=200$, but is obtained with substantially lower computational requirements: intrusive PCE requires \SI{2.8}{\times} less wall-clock time, around \SI{137}{\times} fewer core-hours, and \SI{22.5}{\times} less RAM than SA with $N^s=200$. This highlights that intrusive PCE can accurately determine the optimal objective while using a fraction of the computational resources with respect to SA.

We next assess the accuracy of the first-stage decisions, i.e., the actionable bids submitted to the day-ahead energy market and the aFRR reserve capacity market. Fig.~\ref{fig:runtime_scatters}b--\ref{fig:runtime_scatters}c reports the RMSE of the hourly bid trajectories with respect to the $N^s=2000$ reference, computed per time step and averaged over the \SI{24}{h} horizon. SA exhibits comparatively large errors for small and intermediate scenario sets, with noticeable improvements only at high sample sizes. In contrast, intrusive PCE achieves errors of approximately \SI{0.02}{MW}, lower than those of any tested SA configuration, while requiring substantially fewer computational resources. This indicates that the proposed spectral reformulation captures the optimal first-stage bids with high fidelity.

\begin{figure*}
    \centering
    \includegraphics[width=\textwidth]{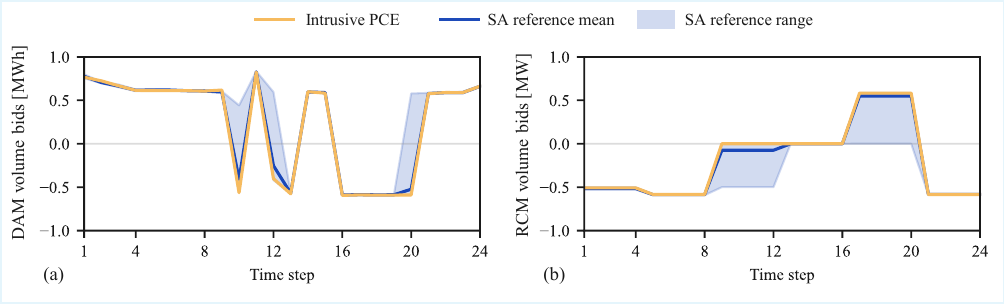}
    \caption{First-stage bid trajectories obtained with intrusive PCE compared with the 2000-scenario SA reference. Solid blue lines denote the reference mean, and the shaded bands indicate the min--max range across the 20 SA repetitions; the yellow lines show the intrusive PCE solution for (a) day-ahead energy (DAM) and (b) reserve-capacity (RCM) volume bids.}
    \label{fig:first_stage_decisions_comparison}
\end{figure*}

Fig.~\ref{fig:first_stage_decisions_comparison} further supports this conclusion. For both the day-ahead energy schedule and the reserve-capacity offers, the intrusive PCE bid trajectories remain closely aligned with the SA reference over the full \SI{24}{h} horizon. In particular, the intrusive PCE solution lies close to the reference mean and remains largely within the min--max envelope across the 20 SA repetitions. The largest deviations occur during time periods when the SA solutions themselves exhibit the greatest variability.

\subsection{Validation of the chance-constraint approximation}
\label{sec:results_oos}
To obtain a tractable formulation in the spectral domain, the original inequality constraints are reformulated as chance constraints and enforced through the moment-based SOC approximation in \eqref{eq:soc_chance_coeff}. This approximation requires selecting a safety factor $\lambda$ a priori, which is typically interpreted under an assumed distribution of the constraint residuals (Gaussian in this study). If this distributional assumption is violated, the true violation probabilities may exceed the target level, potentially leading to an under-enforcement of the original inequalities.

To assess the validity of the approximation, we perform an empirical validation by evaluating $10{,}000$ independent uncertainty realizations for each inequality-constraint residual and estimating empirical violation probabilities. Out of $161{,}613$ inequality constraints, only 31 exhibit an empirical violation share above $5\%$, with a maximum exceedance probability of $5.17\%$. Overall, these results indicate that, for the present uncertainty model and operating conditions, the Gaussian assumption provides an adequate approximation of the residual behavior. Furthermore, the moment-based SOC reformulation enforces the chance constraints with negligible exceedance.

\section{Discussion}
\label{sec:discussion}
\textit{Implications for VPP operators.}
The computational gains observed in the case study are relevant for VPP operators in several aspects. First, the reduced wall-clock time enables day-ahead offers to be computed closer to market gate closure, thereby allowing the use of more recent information on prices and DER availability. Importantly, this is achieved while retaining a stochastic formulation, thus preserving the economic benefits of stochastic optimization over conservative robust counterparts. Moreover, the low runtime and memory requirements improve the practical scalability of the framework with respect to portfolio size, supporting the coordinated scheduling of larger DER fleets.
Second, the proposed approach exhibits favorable scaling with uncertainty dimensionality. For the linear two-stage setting considered here, a first-degree expansion yields a coefficient-space formulation whose size grows linearly with the number of independent uncertainty sources. This supports the incorporation of additional uncertainty drivers without a dimensionality explosion.

\textit{A generalized projection framework for intrusive PCE.}
Beyond the present VPP application, an important contribution of this work is the released projection framework that automates the intrusive PCE reformulation. 
The user provides: 1) the optimization model in deterministic form, i.e., objective function, decision variables, parameters, and constraints; 2) the stage assignment of the decision variables, distinguishing first-stage and recourse variables; 3) the uncertain parameters and their marginal probability distributions; and 4) the projection settings, such as truncation order and chance-constraint calibration parameters. From this input, the framework automatically generates the deterministic coefficient-space counterpart. In particular, it i) constructs the orthogonal polynomial basis and the associated multi-index set, ii) expands the recourse variables in that basis, iii) projects the objective function and equality constraints via the stochastic Galerkin method, and iv) reformulates the inequality constraints via the moment-based SOC approximation. The output is a deterministic optimization model that can be solved directly with standard conic solvers.
This framework reduces the implementation efforts by eliminating the need to manually implement the projection steps. In addition, it improves reproducibility and facilitates the application and benchmarking of intrusive PCE across other single-stage and two-stage stochastic programs with minimal development effort.

\textit{Limitations.}
The proposed framework has several limitations. First, the treatment of inequality constraints is approximate: they are reformulated as chance constraints and then enforced through a moment-based SOC approximation. Although the empirical validation showed negligible exceedance in the present case study, exact equivalence with the original stochastic inequalities can not be guaranteed. Such equivalence depends on the residual distribution and the selected safety factor.
Second, the resulting deterministic counterpart remains a large monolithic SOCP. It is substantially more tractable than the scenario-based benchmark, but it is not trivial to solve. No dedicated decomposition or parallel solution method for the coefficient-space problem has been developed in this work. Third, the present results are obtained for a linear two-stage scheduling model with independent uncertainty sources. Extensions to correlated inputs, nonlinear network or device models, and multistage market formulations remain unexplored.

\section{Conclusions}
\label{sec:conclusions}
This paper proposes an efficient multi-market scheduling framework for DER-based virtual power plants (VPPs) via intrusive Polynomial Chaos Expansion (PCE). The approach reformulates the stochastic program in the spectral domain by expanding the second-stage recourse decisions, projecting equality constraints via stochastic Galerkin projection, and enforcing inequality constraints through a tractable moment-based chance-constraint approximation. The resulting coefficient-space formulation is a deterministic convex conic program that can be solved with standard optimization tools.

The proposed method is demonstrated on a DER-based VPP operating on a realistic Swiss low-voltage grid and benchmarked against a state-of-the-art scenario approximation solved via Benders decomposition. The results show that intrusive PCE achieves solution quality comparable to the scenario-based reference while substantially reducing computational requirements by avoiding scenario enumeration. Specifically, at comparable objective-value accuracy, intrusive PCE requires \SI{2.8}{\times} less wall-clock time, \SI{137}{\times} fewer core-hours, and \SI{22.5}{\times} less RAM than the corresponding scenario-based approximation. In addition, the spectral formulation yields first-stage market decisions that are more accurate than the tested scenario-based configurations while requiring only a fraction of the computational resources. Additionally, the released open-source, application-agnostic projection tool also enhances the practical relevance of this work by enabling intrusive PCE to be applied to generic single- and two-stage stochastic programs with limited implementation effort.

Future work will extend the uncertainty model to capture correlations among forecast errors and will investigate decomposition and parallel solution schemes for the coefficient-space formulation to further enhance scalability.


\bibliographystyle{IEEEtran}
\bibliography{refs}

\end{document}